\documentclass[12pt,a4paper,fleqn]{article}
\usepackage{lineno}
\usepackage{graphicx}
\usepackage{caption}
\usepackage{amsmath}
\usepackage{amssymb}
\usepackage[margin=1in]{geometry}

\begin{document}

\begin{center}
    {\Large \bf A fast radio burst associated with a Galactic magnetar}
\end{center}
\vspace{1cm}
{\large Authors: C. D. Bochenek$^{1,2,3,*}$, V. Ravi$^{2}$, K. V. Belov$^{4}$, G. Hallinan$^{2}$, J. Kocz$^{2,5}$, S. R. Kulkarni$^{2}$ \& D. L. McKenna$^{6}$
} \\

\vspace{0.3cm}

\noindent $^{1}$ Cahill Center for Astronomy and Astrophysics, MC\,249-17, California Institute of Technology, Pasadena CA 91125, USA. \\
$^{2}$ Owens Valley Radio Observatory, MC\,249-17, California Institute of Technology, Pasadena CA 91125, USA. \\
$^{3}$ E-mail: cbochenek@astro.caltech.edu. \\
$^{4}$ Jet Propulsion Laboratory, California Institute of Technology, Pasadena, CA 91109, USA. \\
$^{5}$ Department of Astronomy, University of California, Berkeley, CA 94720, USA. \\
$^{6}$ Caltech Optical Observatories, MC\,249-17, California Institute of Technology, Pasadena CA 91125, USA. \\

\clearpage

\textbf{Since their discovery in 2007, much effort has been devoted to uncovering the sources of the extragalactic, millisecond-duration fast radio bursts (FRBs)$^{1}$. A class of neutron star known as magnetars is a leading candidate source of FRBs. Magnetars have surface magnetic fields in excess of $10^{14}$\,G, the decay of which powers a range of high-energy phenomena$^{2}$. Here we present the discovery of a millisecond-duration radio burst from the Galactic magnetar SGR 1935+2154, with a fluence of $\mathbf{1.5\pm0.3}$ Mega-Jansky milliseconds. This event, termed ST 200428A(=FRB 200428), was detected on 28 April 2020 by the STARE2 radio array$^{3}$ in the 1281--1468\,MHz band. The isotropic-equivalent energy released in ST 200428A is $\mathbf{4\times10^{3}}$ times greater than in any Galactic radio burst previously observed on similar timescales. ST 200428A is just 40 times less energetic than the weakest extragalactic FRB observed to date$^{4}$, and is arguably drawn from the same population as the observed FRB sample. The coincidence of ST 200428A with an X-ray burst$^{5,6,7}$ favours emission models developed for FRBs that describe synchrotron masers or electromagnetic pulses powered by magnetar bursts and giant flares$^{8,9,10,11}$. The discovery of ST 200428A implies that active magnetars like SGR 1935+2154 can produce FRBs at extragalactic distances. The high volumetric rate of events like ST 200428A motivates dedicated searches for similar bursts from nearby galaxies.}


\vspace{0.5cm}
{\bf Discovery of ST 200428A.} 
Three 1281--1468\,MHz radio detectors comprise STARE2$^{3}$, and are located across the south-western United States. All three detectors were triggered by ST 200428A at an Earth-centre arrival time at infinite frequency of UTC 28 April 2020 14:34:24.45481(3). We hereafter quote the standard errors in the last significant figures in parentheses. The detected signal-to-noise ratios of ST 200428A were 21, 15 and 20 (see Methods). The burst was detected with a dispersion measure (DM) of 332.702(8)\,pc\,cm$^{-3}$ and an intrinsic width of 0.61(9)\,ms (Figure~1). The band-averaged fluence of ST 200428A was $1.5(3)\times10^{6}$\,Jy\,ms. See Table~1 for other properties.



On 27 April 2020, the Swift Burst Alert Telescope reported multiple bursts from the soft $\gamma$-ray repeater (SGR) 1935+2154 [15], indicating that SGR 1935+2154 had entered a phase of heightened activity. One day later, the CHIME/FRB collaboration first reported an approximately $10^{3}$\,Jy\,ms burst from 400\,MHz--800\,MHz from the approximate direction of SGR 1935+2154, which was outside the direct field of view of CHIME at the time (Figure~2)$^{16}$. We responded by expediting our daily manual inspection of STARE2 triggers recorded throughout the day, and found that ST 200428A was detected at approximately the same time and DM as the CHIME event, but with a roughly one thousand times higher fluence. The localisation region of ST 200428A includes SGR 1935+2154 (Figure~2).
Shortly thereafter, a constellation of space-borne instruments$^{5,6,7}$ 
reported a one-second-long X-ray (1--250\,keV) burst from the direction of SGR 1935+2154 that occurred at precisely the same time as the CHIME bursts and ST 200428A (Figure~1). Two days later, on 30 April 2020, the Five hundred meter Aperture Spherical Telescope (FAST) reported a weak (0.06\,Jy\,ms) radio pulse from the direction (within 3\,arcminutes) of SGR 1935+2154 [17], with a DM consistent with the CHIME and STARE2 events. The coincident DMs between the CHIME, STARE2 and FAST detections, combined with the accurate FAST localisation, establish a unique source of dispersed radio bursts in the direction of SGR 1935+2154. The temporal coincidence between the CHIME, STARE2 and X-ray detections on 28 April 2020 identify them as a singular burst from SGR 1935+2154.

{\bf Observational classification.} 
The fluence and isotropic-equivalent energy release of ST 200428A is extremely unusual among the known Galactic radio emitters (Figure~3). We adopt a distance to SGR 1935+2154 of 9.5\,kpc, which is consistent with analyses of high-energy flares from the SGR$^{22}$, and in the middle of the range of distances estimated for the supernova remnant associated with the SGR$^{18,19}$. 
The isotropic-equivalent energy release of ST 200428A was $2.2(4)\times10^{35}$\,erg, and the spectral energy release was $1.6(3)\times10^{26}$\,erg\,Hz$^{-1}$ (see Methods).  Previously, giant radio pulses (GRPs) from a small subset of young and millisecond pulsars formed the most energetic class of sub-millisecond emission within the Milky Way at centimetre wavelengths$^{23}$. GRPs are emitted stochastically amongst regular radio pulses, and have sub-microsecond durations (unlike ST 200428A). The brightest and most luminous GRP reported in the literature was observed at 430\,MHz from the Crab pulsar, with a fluence of approximately $3\times10^{4}$\,Jy\,ms and an isotropic-equivalent energy release of $6\times10^{31}$\,erg [24]. ST 200428A is therefore a factor of approximately $4\times10^{3}$ more energetic than any millisecond radio burst previously observed from a source within the Milky Way. 


From Figure~3, we see that ST 200428A is most plausibly related to the fast radio bursts (FRBs) observed at extragalactic distances. FRBs span a wide range of energies, but the least energetic FRB reported to date (from FRB\,180916.J0158+65) had an isotropic-equivalent spectral energy of just $5(2) \times 10^{27}$\,erg\,Hz$^{-1}$  at 1.7\,GHz [4]. The brightness temperature of ST 200428A, assuming a duration of 0.61\,ms and a mean flux density of 2.5\,MJy, is approximately $1.4\times10^{32}$\,K, which is consistent with the $10^{32}-10^{37}$\,K brightness temperatures of FRBs$^{1}$. The factor of $10^{3}$ disparity between the fluence of ST 200428A and the fluence of the temporally coincident lower-frequency burst detected by CHIME is consistent with previous wide-band radio observations of FRBs$^{25}$. It is likely that current FRB surveys are incomplete at low fluences$^{26}$. This implies that a substantial population of sub-threshold events likely exists that could span the small energy gap between known FRBs and ST 200428A.

STARE2 had been observing for 448 days prior to ST 200428A, with an effective field of view of 1.84 steradian and a detection threshold of 300\,kJy for millisecond duration bursts (see Methods). Based on the entirety of the STARE2 observing campaign, we derive a volumetric rate for bursts with energy releases equivalent to or greater than ST 200428A of $7.23_{-6.13}^{+8.78} \times 10^7$\,Gpc$^{-3}$\,yr$^{-1}$. This is consistent with an extrapolation of the luminosity function of bright FRBs (see Methods).

SGR 1935+2154 is located approximately 100\,pc above the plane of the Milky Way disk and approximately 9\,kpc from the Galactic centre, at the centre of a known supernova remnant$^{18}$. 
This locale is consistent with those of four of the five accurately localised FRBs within their host galaxies$^{27,28,29,4}$. These four host galaxies are similar to the Milky Way in their masses and star-formation rates. Furthermore, the gaseous environments of these FRBs are similar to that of SGR 1935+2154. There is no evidence for a significant DM enhancement locally to SGR 1935+2154 (see Methods). The Faraday rotation measure (RM) of the weak radio pulse detected by FAST from SGR 1935+2154 is consistent with the RM of radio emission from the associated supernova remnant at the position of SGR 1935+2154 [18], implying no significant enhancement in the plasma magnetisation immediately surrounding the SGR. 

{\bf Implications for FRBs. }
We have established that ST 200428A is arguably drawn from the same population as the observed FRB sample. This implies that magnetars like SGR 1935+2154 are viable FRB engines. However, it is puzzling that events like ST 200428A have not previously been observed from SGR 1935+2154, or from any of the 30 known magnetars in the Milky Way$^{2}$. The rotation period (3.24\,s) and surface dipole magnetic field ($2.2\times10^{14}$\,G) of SGR 1935+2154 is typical of the Milky Way magnetar population$^{2}$. Although SGR 1935+2154 was in a phase of increased X-ray burst activity when ST 200428A was detected, increased burst activity has previously been observed from this SGR$^{30}$. Indeed, SGR 1935+2154 is the most prolific known burster among the magnetar population$^{30}$. The X-ray burst coincident with ST 200428A (isotropic-equivalent energy release of $8.3(8)\times10^{39}$\,erg) was a typical example of a magnetar burst$^{2}$, with perhaps some unusual spectral characteristics$^{6}$. The FAST detection of a weak radio burst from SGR 1935+2154 [17] is consistent with previous observations of pulsed radio emission from four Milky Way magnetars$^{2}$. 


The properties of ST 200428A and the rarity of similar events provide insight into the emission mechanism of FRBs. The temporal coincidence of ST 200428A with an X-ray burst is not fully consistent with models in which FRB emission is generated within the magnetospheres of magnetars. If the X-ray burst is emitted through standard mechanisms, magnetospheric FRB emission is predicted to occur immediately prior to the X-ray bursts$^{31}$. Indeed, X-ray bursts appear to suppress magnetospheric radio emission from the magnetar-like pulsar PSR\,J1119-6127 for several tens of seconds$^{32}$.

The energetics of ST 200428A and its coincident X-ray burst are consistent with some models for FRB emission from beyond the magnetospheres of magnetars.
Models for FRB emission external to magnetospheres have only been developed for the most energetic X- and $\gamma$-ray bursts from magnetars, known as giant flares.  An event that manifests as a giant flare is also predicted to result in the ejection of a highly magnetised portion of the magnetar magnetosphere at relativistic speeds, known as a ``plasmoid''. Although plasmoids are distinct from the observed X- and $\gamma$-ray radiation from giant flares, they are expected to contain comparable amounts of energy  ($>10^{44}$\,erg)$^{8,9,10}$. 
An FRB radiated by the synchrotron maser mechanism may result from a shock driven by the plasmoid into the external medium$^{8,9,10}$. Alternatively, the plasmoid may trigger an electromagnetic pulse just beyond the magnetosphere, resulting in an FRB$^{11}$. FRB energies of $10^{-7}E_{\rm pl}$ to $10^{-4}E_{\rm pl}$ are predicted by these models. The ratio between the radio energy release of ST 200428A and the X-ray energy release of the coincident burst is approximately $3\times10^{-5}$, consistent with these predictions for the energy of the radio burst. 

In this scenario, individual X-ray bursts from magnetars would result in radio emission being relativistically beamed in random directions, explaining the rarity of events like ST 200428A. However, it remains to be seen whether the theory developed for plasmoids launched by giant magnetar flares can be extended to lower-energy events like the $10^{40}$\,erg X-ray burst coincident with ST 200428A. The close coincidence between ST 200428A and the associated X-ray burst (Figure 1) is difficult to explain if the FRB were launched extremely far from the magnetosphere$^{8}$, unless the X-ray emission was not associated with the FRB-emitting region.

Our observations suggest that magnetars like those observed in the Milky Way can produce FRBs. The consistency between the volumetric rate of events like ST 200428A and the extrapolated FRB volumetric rate suggests that we have observed the dominant FRB channel, if all FRBs repeat$^{26}$. Magnetars have been proposed to be born in core-collapse supernovae, the accretion induced collapse of white dwarfs, binary white dwarf mergers, mergers of white dwarfs with neutron stars, and binary neutron star mergers. No evidence exists within the Milky Way for any magnetar formation channel besides core-collapse supernovae. Observations of the host galaxies and environments of extragalactic FRBs may reveal the full range of magnetar formation channels. For example, an FRB observed from a galaxy lacking ongoing star formation would immediately suggest a previously unobserved magnetar formation channel. 


ST 200428A is a new class of radio emission from within the Milky Way (Figure 3), and evidence suggests that it is a Galactic analogue of the extragalactic FRBs. The rapid detection of ST 200428A by STARE2 implies a high volumetric rate of similar bursts (see Methods). Its association with a young (age of approximately $10^{4}$\,yr)$^{2,18}$ magnetar formed in a core-collapse supernova implies an association of similar bursts with ongoing star-formation activity. Our observations motivate dedicated searches for events like ST 200428A from nearby, rapidly star-forming galaxies like Messier 82, which has a star-formation rate approximately 40 times that of the Milky Way$^{3}$. 

\clearpage

\noindent {\bf References}

\begin{enumerate}
    
    \item Petroff, E., Hessels, J. W. T. \& Lorimer, D. R. Fast radio bursts. {\em Astron. Astrophys. Rev.} {\bf 27,} 4 (2019).
    \item Kaspi, V. M. \& Beloborodov, A. M. Magnetars. {\em Annu. Rev. Astron. Astrophys.} {\bf 55,} 261 (2017).
    \item Bochenek, C. D. {\em et al.} STARE2: Detecting Fast Radio Bursts in the Milky Way. {\em Publ. Astron. Soc. Pac.} {\bf 132,} 034202 (2020).
    \item Marcote, B. {\em et al.} A repeating fast radio burst source localized to a nearby spiral galaxy. {\em Nature} {\bf 577,} 190 (2020). 
    \item Mereghetti, S. {\em et al.} SGR 1935+2154: INTEGRAL hard X-ray counterpart of radio burst. {\em GRB Circular Network} {\bf 27668,} 1 (2020).
    \item Ridnaia, A., {\em et al.} Konus-Wind observation of hard X-ray counterpart of the radio burst from SGR 1935+2154. {\em GRB Circular Network} {\bf 27669,} 1 (2020). 
    \item Zhang, S.-N. {\em et al.} Insight-HXMT detection of a bright short x-ray counterpart of the Fast Radio Burst from SGR 1935+2154. {\em The Astronomer's Telegram} {\bf 13687,} 1 (2020).
    \item Lyubarsky, Y. A model for fast extragalactic radio bursts. {\em Mon. Not. R. Astron. Soc.} {\bf 442,} L9 (2014). 
    \item Beloborodov, A. M. A Flaring Magnetar in FRB 121102? {\em Astrophys. J. Lett.} {\bf 843,} L26 (2017).
    \item Metzger, B. D., Margalit, B. \& Sironi, L. Fast radio bursts as synchrotron maser emission from decelerating relativistic blast waves. {\em Mon. Not. R. Astron. Soc.} {\bf 485,} 4091 (2019).
    \item Lyubarsky, Y. Fast radio bursts from reconnection in magnetar magnetosphere. Preprint at \\ https://arxiv.org/abs/2001.02007 (2020). 
    \item Zhang, S.-N. {\em et al.} Insight-HXMT X-ray and hard X-ray detection of the double peaks of the Fast Radio Burst from SGR 1935+2154. {\em The Astronomer's Telegram} {\bf 13696,} 1 (2020). 
    \item Zhang, S.-N. {\em et al.} Geocentric time correction for Insight-HXMT detection of the x-ray counterpart of the FRB by CHIME and STARE2 from SGR 1935+2154. {\em The Astronomer's Telegram} {\bf 13704,} 1 (2020). 
    \item Manchester, R. N. \& Taylor, J. H. {\em Pulsars.} (W. H. Freeman, 1977).
    \item Barthelmy, S. D. {\em et al.} Swift detection of multiple bursts from SGR 1935+2154. {\em GRB Circular Network} {\bf 27657,} 1 (2020).
    \item Scholz, P. {\em et al.} A bright millisecond-timescale radio burst from the direction of the Galactic magnetar SGR 1935+2154. {\em The Astronomer's Telegram} {\bf 13681,} 1 (2020).
    \item Zhang, C. F. {\em et al.} A highly polarised radio burst detected from SGR 1935+2154 by FAST. {\em The Astronomer's Telegram} {\bf 13699,} 1 (2020). 
    \item  Kothes, R., Sun, X., Gaensler, B. \& Reich, W. A Radio Continuum and Polarization Study of SNR G57.2+0.8 Associated with Magnetar SGR 1935+2154. {\em Astrophys. J.} {\bf 852,} 54 (2018). 
    \item Zhou, P. {\em et al.} Revisiting the distance, environment and supernova properties of SNR G57.2+0.8 that hosts SGR 1935+2154. Preprint at \\ https://arxiv.org/abs/2005.03517 (2020). 
    \item Keane E. F. The future of fast radio burst science. {\em Nature Astronomy} {\bf 2,} 865 (2018).
    \item Villadsen, J. \& Hallinan, G. Ultra-wideband Detection of 22 Coherent Radio Bursts on M Dwarfs. {\em Astrophys. J.} {\bf 871,} 214 (2019). 
    \item Kozlova, A. V. {\em et al.} The first observation of an intermediate flare from SGR 1935+2154. {\em Mon. Not. R. Astron. Soc.} {\bf 460,} 2008 (2016). 
    \item Kuzmin, A. D. Giant pulses of pulsar radio emission. {\em Astrophys. Space Sci.} {\bf 308,} 563 (2007).
    \item Cordes, J. M., Bhat, N. D. R., Hankins, T. H., McLaughlin, M. A. \& Kern, J. The Brightest Pulses in the Universe: Multifrequency Observations of the Crab Pulsar's Giant Pulses. {\em Astrophys. J.} {\bf 612,} 375 (2004). 
    \item Sokolowski, M. {\em et al.} No Low-frequency Emission from Extremely Bright Fast Radio Bursts. {\em Astrophys. J.} {\bf 867,} L12 (2018). 
    \item Ravi, V. The prevalence of repeating fast radio bursts. {\em Nature Astronomy} {\bf 3,} 928 (2019). 
    \item Bannister, K. {\em et al.} A single fast radio burst localized to a massive galaxy at cosmological distance. {\em Science} {\bf 365,} 565 (2019).
    \item Ravi, V. {\em et al.} A fast radio burst localized to a massive galaxy. {\em Nature} {\bf 572,} 352 (2019). 
    \item Prochaska, J. X. {\em et al.} The low density and magnetization of a massive galaxy halo exposed by a fast radio burst. {\em Science} {\bf 366,} 231 (2019).
    \item Lin, L. {\em et al.} Burst Properties of the Most Recurring Transient Magnetar SGR J1935+2154. {\em Astrophys. J.} {\bf 893,} 156 (2020). 
    \item Lu, W. \& Kumar, P. On the radiation mechanism of repeating fast radio bursts. {\em Mon. Not. R. Astron. Soc.} {\bf 477,} 2470 (2018).
    \item Archibald, R. F., Kaspi, V. M., Tendulkar, S. P. \& Scholz, P. The 2016 Outburst of PSR J1119-6127: Cooling and a Spin-down-dominated Glitch. {\em Astrophys. J.} {\bf 869,} 180 (2018).

\end{enumerate}

\vspace{1cm}

\noindent {\bf Supplementary Information} is linked to the online version of the paper at \\ www.nature.com/nature.

\vspace{0.5cm}

\noindent {\bf Acknowledgements.} We would like to thank the then director of OVRO, A. Readhead, for funds (derived from the Alan Moffet Funds) which allowed us to start this project. The Caltech and Jet Propulsion Laboratory President's and Director’s Fund allowed us to build the
second system at Goldstone and third system near Delta, Utah. We are thankful to Caltech
and the Jet Propulsion Laboratories for the second round of funding. Christopher Bochenek, a PhD student, was partially supported by the Heising-Simons foundation. We would also like to thank Sandy Weinreb and David Hodge for building the front end and back end boxes, Jeffrey Lagrange for his support at GDSCC, John Matthews and the Telescope Array collaboration for their support at Delta, and the entire OVRO staff for their support,  in particular James Lamb, David Woody, and Morgan Catha. We thank Sterl Phinney and Wenbin Lu for comments on the manuscript. A portion of this research was performed at the Jet Propulsion Laboratory, California Institute of Technology, under a contract with the National Aeronautics and Space Administration. This research was additionally supported by the National Science Foundation under grant AST-1836018. This research made use of Astropy, a community-developed core Python package for Astronomy. This research has also made use of the SIMBAD database, operated at CDS, Strasbourg, France.

\vspace{0.5cm}

\noindent {\bf Author contributions.}
S.R.K., C.D.B., D.L.M., V.R., K.V.B., and G.H.~conceived of and developed the STARE2 concept and observing strategy. C.D.B., D.L.M., K.V.B., J.K., and S.R.K.~led the construction and initial deployment of STARE2. C.D.B., D.L.M., K.V.B., V.R., J.K., and G.H.~designed and built the STARE2 subsystems. C.D.B., D.L.M., and K.V.B.~commissioned STARE2. C.D.B.~operated STARE2, performed the localisation and transient rate analyses, as well as the searches for sub-threshold events and events associated with other SGR flares. V.R.~extracted the properties of the burst. C.D.B.~and V.R.~led the writing of the manuscript with the assistance of all co-authors.

\vspace{0.5cm}

\noindent {\bf Author information.} Reprints and permissions information is available at \\ www.nature.com/reprints. The authors declare no competing interests. Correspondence and requests for materials should be addressed to cbochenek@astro.caltech.edu. 

\clearpage

\begin{figure}
    \centering
    \includegraphics[width=0.7\textwidth]{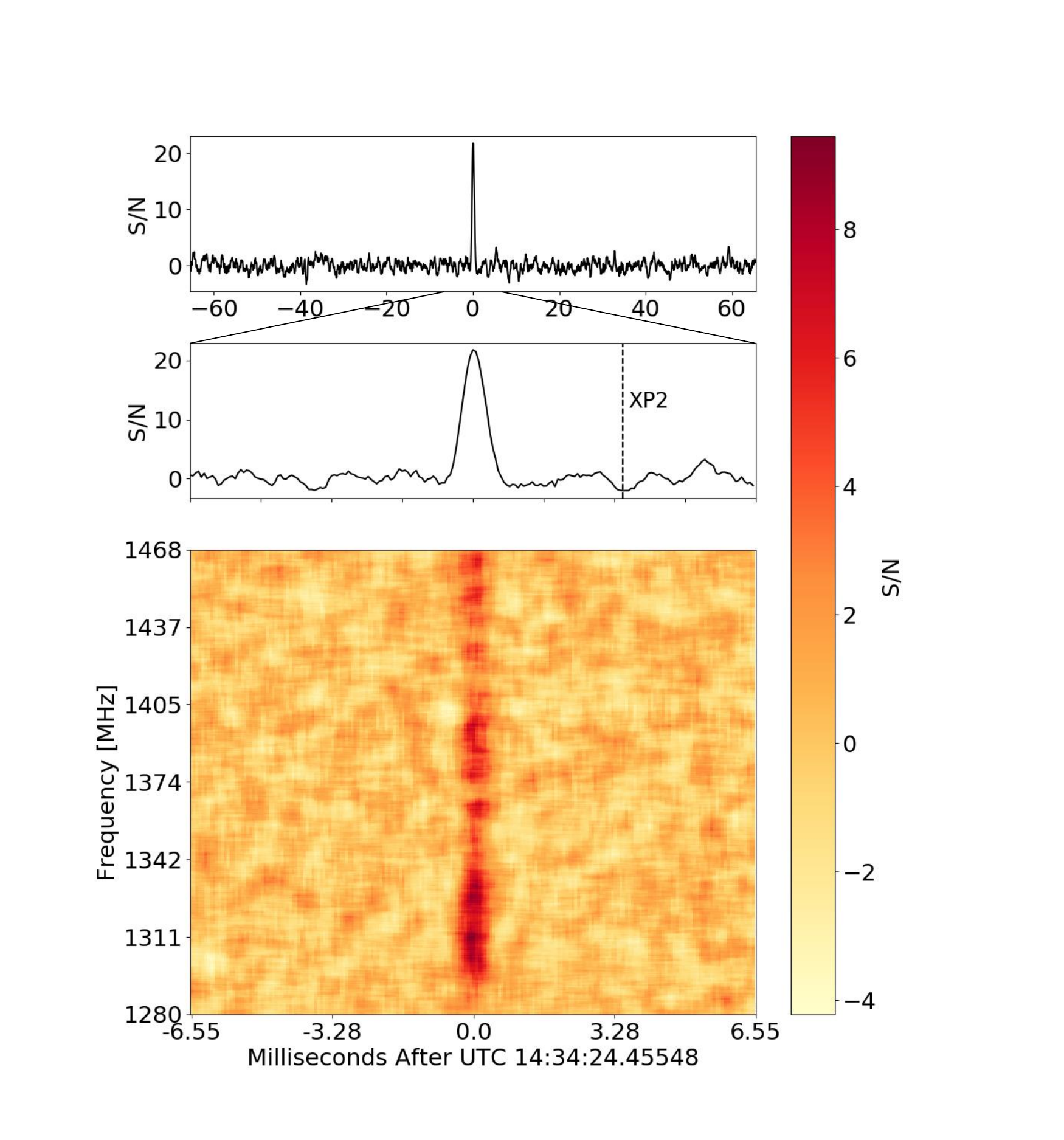}
    \caption{{\bf Time-series and dynamic spectrum of ST 200428A.} 
    We show data obtained from the Owens Valley Radio Observatory (OVRO) alone. All data units are signal to noise ratio (S/N). The quoted times are relative to the Earth-centre arrival time of the burst at infinite frequency.
    For a description of the data processing, see the Methods section. \textit{Top:} De-dispersed time series of all available data on ST 200428A (see Methods). The original data were de-dispersed at a DM of 332.702\,pc\,cm$^{-3}$. We detect no other radio bursts within our data, spanning a window of 131.072\,ms centred on the time of ST 200428A. We place an upper limit on bursts with S/N$>5$ in this time window of $<400$\,kJy\,ms. \textit{Middle:} Expanded plot of the region surrounding the burst. The relative arrival time of the second, brighter peak in the coincident X-ray burst (XP2) is indicated as a vertical dashed line$^{12,13}$. The full X-ray burst lasted approximately 1\,s centred on UTC 14:34:24.444 (arrival time at the Earth center).  \textit{Bottom:} Dynamic spectrum of ST 200428A corrected for the effects of dispersion.} 
    \label{fig:waterfall}
\end{figure}

\begin{figure}
    \centering
    \includegraphics[width=\textwidth]{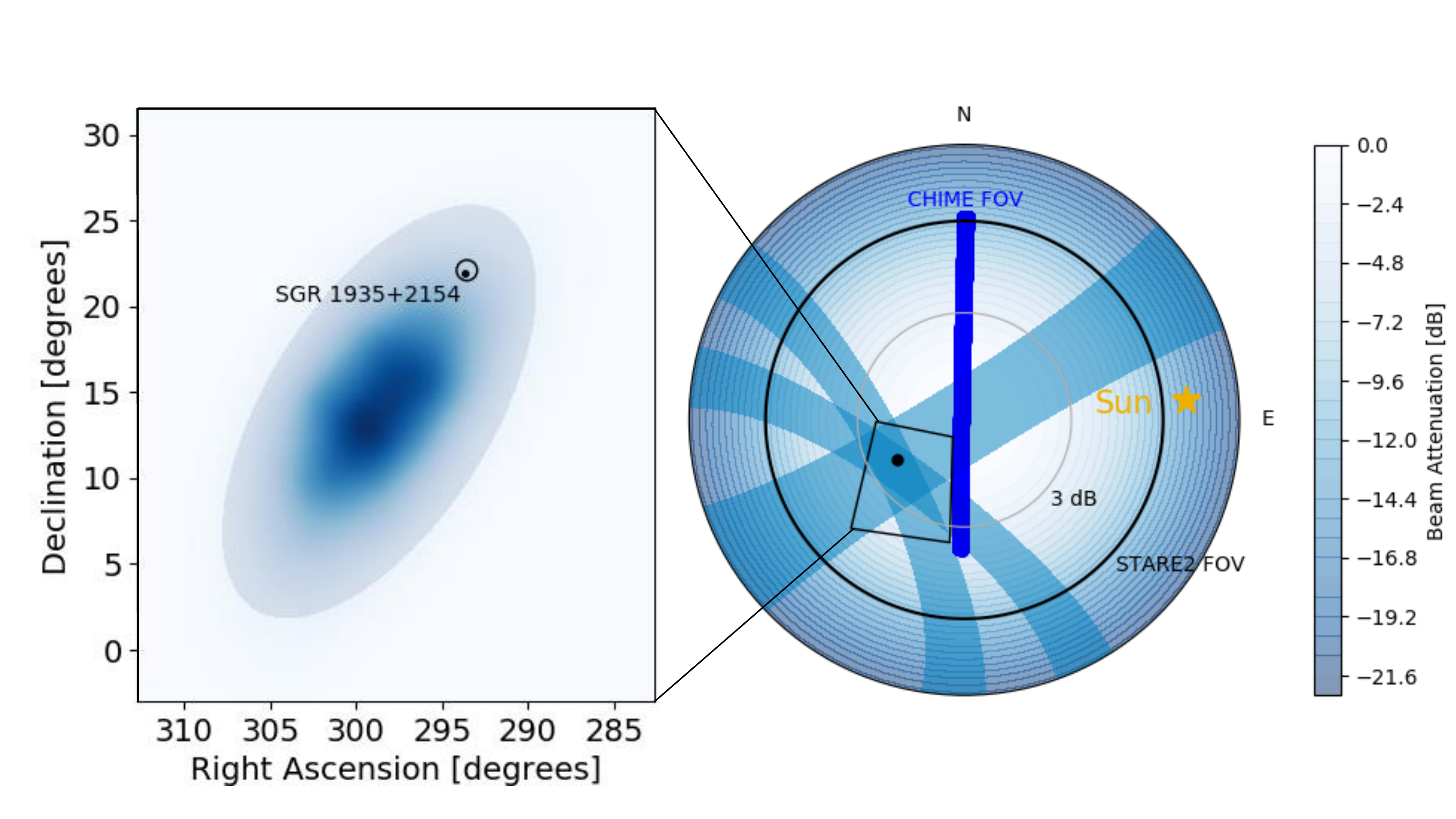}
    \caption{{\bf STARE2 localisation of ST 200428A.} \textit{Right:} An altitude and azimuth view of the sky at the OVRO STARE2 station at the time ST 200428A was detected. The large black circle corresponds to the STARE2 field of view (FOV), which is set by the edge of a horizon shield at OVRO$^{3}$. A grey circle labelled ``3 dB'' indicates the zenith angle corresponding to the FWHM of the STARE2 response on the sky. The thick blue line represents the CHIME FOV. The yellow star represents the Sun, which is a common source of STARE2 triggers$^{3}$. The black dot represents the known position of SGR 1935+2154. The three light blue arcs correspond to the 95\%-confidence localisations for each individual STARE2 baseline. The black quadrilateral represents the outline of the region shown in the left panel. \textit{Left:} The 95\% confidence STARE2 localisation region of ST 200428A is shown as a blue ellipse. The blue gradient corresponds to the probability the burst occurred at that location. The CHIME localisation region$^{16}$ corresponds approximately to the black circle. The known position of SGR 1935+2154, which is identical to the position of the weak burst detected by FAST$^{17}$, is shown as a black dot.}
    \label{fig:loc}
\end{figure}

\begin{figure}
    \centering
    \includegraphics[width=0.9\textwidth,trim=1cm 0cm 7cm 0cm, clip]{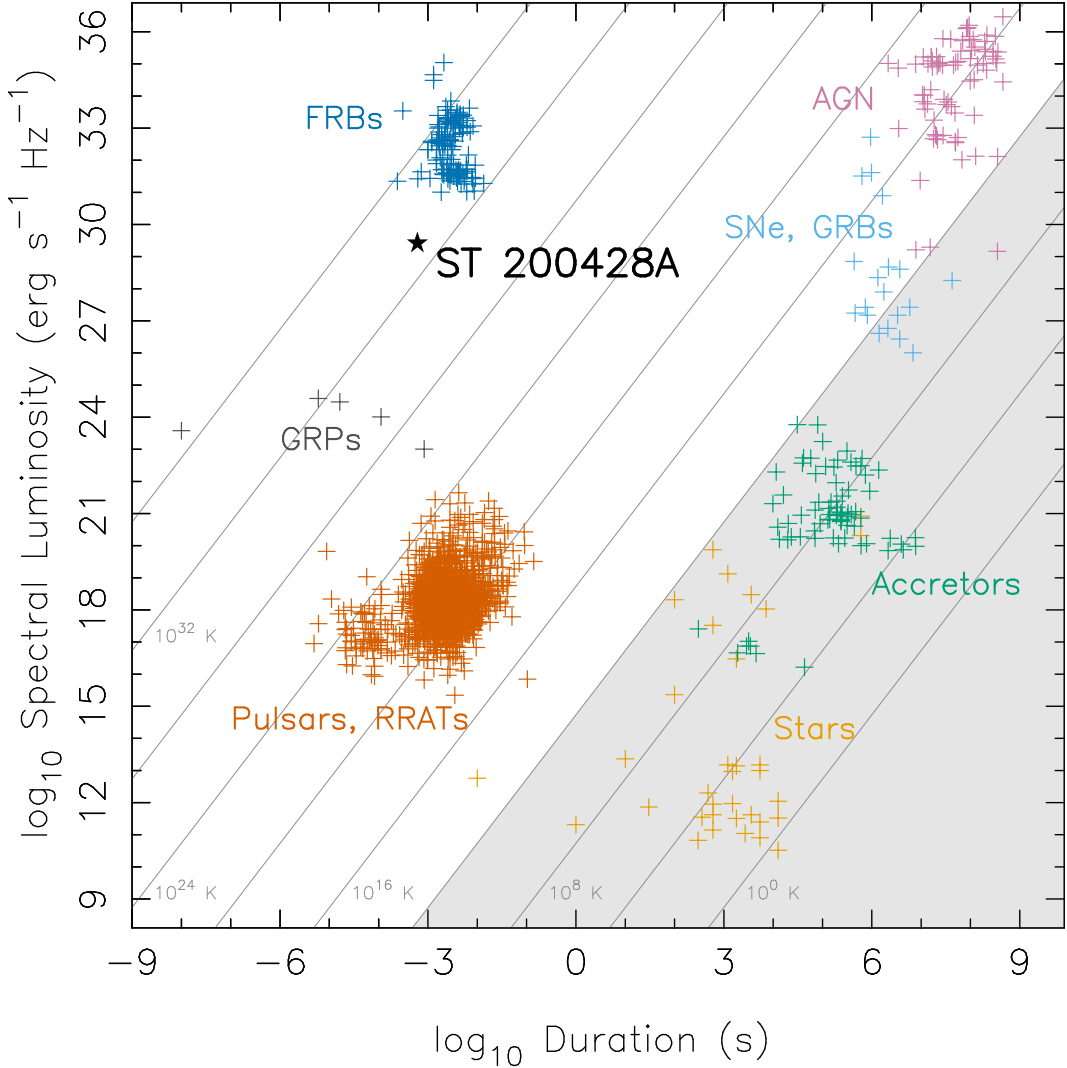}
    \caption{{\bf Phase space of centimetre-wavelength radio transient events.} The vertical extent of the ST 200428A star corresponds to the uncertainty in the spectral luminosity caused by the uncertain distance to SGR 1935+2154, which ranges between 6.5--12.5\,kpc [18,19]. Only isotropic-equivalent spectral luminosities are shown. The FRBs plotted only include bursts detected between 1--2\,GHz from sources at known distances. All other data were gathered from Refs. 20 and 21. ``GRPs'' refers to giant radio pulses, ``RRATs'' refers to rotating radio transients, ``accretors'' refers to accreting binary systems in the Milky Way, ``SNe'' and ``GRBs'' refers to supernovae and $\gamma$-ray bursts at extragalactic distances, and ``AGN'' refers to accreting supermassive black holes. Lines of constant brightness temperature at a reference frequency of 1.4\,GHz are shown, and the shaded area (representing brightness temperatures less than $10^{12}$\,K) indicates sources that are likely incoherent emitters that are not relativistically boosted. The spectral luminosity of ST 200428A was derived by dividing the burst spectral energy by the  duration.}  
    \label{fig:phase_space}
\end{figure}

\clearpage


\begin{table}
	\centering
	\caption{{\bf Data on ST 200428A.} Standard errors in the final significant figures (68\% confidence) given in parentheses. \\
	$^{\rm a}$ The correction to the infinite-frequency ($\nu=\infty$) arrival time is done using the DM quoted in this table, and assuming a dispersion constant of $\frac{1}{2.41}\times10^{4}$\,s\,MHz$^{2}$\,pc$^{-1}$\,cm$^{3}$ [14]. \\
	$^{\rm b}$ The full-width half-maximum (FWHM) of the Gaussian used to model the intrinsic burst structure (Methods). \\
	$^{\rm c}$ This assumes a distance to SGR 1935+2154 of 9.5\,kpc.}
	\label{tab:1}
	\begin{tabular}{ll} 
		\hline
		{\bf Property} & {\bf Measurement} \\
		\hline
		OVRO arrival time at $\nu=1529.267578$\,MHz (UTC) & 28 April 2020 14:34:25.02657(2) \\
		OVRO arrival time at $\nu=\infty^{\rm a}$ (UTC) & 28 April 2020 14:34:24.43627(3) \\
		Earth centre arrival time at $\nu=\infty^{\rm a}$ (UTC) & 28 April 2020 14:34:24.45548(3)  \\
		Fluence (MJy\,ms) & $1.5(3)$ \\
		Dispersion measure (pc\,cm$^{-3}$) & 332.702(8) \\
		Intrinsic burst FWHM$^{\rm b}$ (ms) & 0.61(9) \\
		Isotropic-equivalent energy release$^{\rm c}$ (erg) & $2.2(4) \times10^{35}$ \\
		\hline
	\end{tabular}
\end{table}

\clearpage

\begin{center}
    {\Large \bf Methods}
\end{center}
\vspace{0.5cm}

\noindent {\bf The STARE2 instrument.} STARE2 consists of a set of three radio receivers at locations across the south-western United States. The receivers are located at the Owens Valley Radio Observatory (OVRO), the Goldstone Deep Space Communications Complex (GDSCC), and a site operated by the Telescope Array project$^{33}$ near Delta, Utah. 
Although the STARE2 receivers operate in the 1280\,MHz--1530\,MHz band, the useful band is limited to 1280.732422\,MHz to 1468.232422\,MHz by radio-frequency interference. 
STARE2 is sensitive to fast, dispersed radio transients above a fluence detection threshold of 300\,kJy\,$\sqrt{{\rm ms}}$. STARE2 has a time resolution of 65.536\,$\mu$s and frequency resolution of 122.07 kHz. The instrument and data analysis are further described elsewhere$^{3}$. Since the publication of Ref. 3, our station in Delta, Utah commenced operations. With the addition of the third station, we now visually inspect candidate bursts if any pair of stations identifies a candidate event within 100\,ms of each other.

STARE2 data consist of (a) candidate dispersed bursts that triggered the automated pipeline software, and (b) spectra recorded every 65.536\,$\mu$s with 2048 frequency channels between 1280\,MHz--1530\,MHz. The latter data span the full duration of the dispersion sweep of each candidate burst, with an additional 1000 spectra both before and after the start and end of the dispersion sweep respectively. Such time-series of spectra are referred to as dynamic spectra. After de-dispersion, the available time-series data on each burst span 131.072\,ms, as displayed in Figure 1 for ST 200428A. 

To produce Figure 1, the OVRO dynamic spectrum was smoothed with a 2D boxcar function with a width of 0.524288 ms in time and 7.8125 MHz in frequency. Data from the other stations were not summed for the purposes of display because of the different instrumental spectral responses, but are shown in Extended Data Figure 1. The data were baselined by subtracting the mean of the off-pulse region. We normalised the time series in each frequency bin by dividing by the standard deviation of the time series in each frequency bin. The data were again normalised by the standard deviation of the off-pulse time and frequency bins to measure the signal to noise (S/N) in each time and frequency bin. To produce the time series, we simply took our dynamic spectra, which was processed as described above, and averaged the data in frequency and re-normalised by the standard deviation of the time series in the off-pulse region.

\noindent {\bf Localisation.} To localise ST 200428A, we first measured the relative arrival times of the burst between each station. For this analysis, we used the frequency averaged, $65.536$\,$\mu$s resolution, total intensity time series data. The data were dedispersed, and the time series was convolved, with the dispersion measure (DM) and boxcar function that respectively maximised the signal to noise ratio (S/N) at OVRO. These values are derived differently than the values reported in the main text and thus differ slightly. The DM quantifies the observed frequency-dependent dispersion delay in terms of a free-electron column density. This DM and boxcar width were 333.3\,pc\,cm$^{-3}$ and 0.524288\,ms, respectively. To measure the S/N for this analysis, we take the frequency averaged time series that has been dedispersed and convolved with a boxcar function and subtract the mean of the off-pulse region. After this subtraction, the data are divided by the standard deviation of the time series in the off-pulse region. We measure a S/N at OVRO of 21.6, at GDSCC of 15.7, and at Delta of 20.1. Data from each station are shown in Extended Data Figure 1. 

After determining the S/N at each station and processing the data as described above, we then cross-correlate the time series from each pair of stations. We fit the peak of the correlation curve with a Lorentzian function. The location of the peak of the Lorentzian corresponds to the time delay between the two stations in the baseline. We find a time delay for OVRO-GDSCC of $-7\,\mu$s, $-884\,\mu$s for OVRO-Delta, and $-888\,\mu$s for GDSCC-Delta. The statistical uncertainty in the time delays is given by the boxcar width divided by the S/N of the station with the lowest S/N. For OVRO-GDSCC and GDSCC-Delta, this uncertainty is 33\,$\mu$s and 26\,$\mu$\,s for OVRO-Delta.

To estimate the systematic uncertainty, we take advantage of a test of the Global Positioning System's (GPS) L3 signal at 1381\,MHz on 2019 February 28. We recorded data with a time resolution of $131.072$\,$\mu$s during the test. As the data were taken before the Delta station was built, this analysis was done only with OVRO and GDSCC. However, as we use an identical receiver and GPS timing hardware at the Delta station, we expect similar systematics to be present. During testing, the L3 signal turns on and off, allowing for a test of the measured time delay between each station. The intrinsic time delay of the emission of the L3 signal between two satellites is expected to be of order $\mu$s, as the GPS satellites are synchronised in transmission. Furthermore, the received signal will be dominated by the satellites closest to zenith. At the time of the signal considered, the satellite expected to dominate the signal had an expected time delay of $46$\,$\mu$s between OVRO and GDSCC. This is less than the time resolution of our data. We measure the time delay as described above, except in this analysis we only consider frequencies corresponding to the L3 signal and do not convolve our time series with a boxcar function. We find a systematic uncertainty of $81$\,$\mu$s, which represents the measured time delay of the GPS L3 signal between OVRO and GDSCC.

To turn the measured time delays for ST 200428A into a sky position, for each baseline we calculate the expected time delay for a given sky location over a fine grid in azimuth and elevation at a reference location. The reference location is chosen to be OVRO for the OVRO-GDSCC and OVRO-Delta baselines, while GDSCC is chosen for the GDSCC-Delta. We then transform the grid of azimuths and elevations into right ascensions and declinations for the time of the burst. The localisation for one baseline is then those sky positions that are consistent with the measured time delays and uncertainties. This corresponds to an arc across the sky for one baseline. 

To combine the baselines, we assign a probability to each sky location for each arc assuming Gaussian statistical and systematic uncertainties, parameterised by the distance from the arc with no uncertainty. The mean of the Gaussian corresponds to the arc with no uncertainty and the standard deviation corresponds to the width of the arc assuming a $1\sigma$ uncertainty in the time delay. The probabilities for each arc and sky location were then multiplied together and normalised. We smoothed the probability distribution with a two-dimensional Gaussian of standard deviation 1$^{\circ}$. The transformation between local coordinate systems and the celestial coordinate system produced a sparse array of probabilities that required smoothing to visualise. The smoothing radius is much smaller than the size of the localisation region. This produced the probabilities shown in the right panel of Figure~2 as arcs. 

With these probabilities, we measured the 95\% confidence interval of the STARE2 three-station localisation region by modelling it as an ellipse. The 95\% confidence interval corresponds to the smallest ellipse that encloses 95\% of the probability distribution. To estimate this ellipse, we first measured the orientation of the ellipse. We measured the angle of the semi-major axis with respect to the declination axis by modelling the sky location probability distribution using a principle component analysis with two components. The angle of the semi-major axis corresponds to the angle of the eigenvector of the principle component with the highest eigenvalue. This angle is 57.88$^{\circ}$. We calculated the expectation values of the right ascension and declination directly from their marginalised probability distributions. We find a right ascension of $\alpha = $19d55m$\pm 15^{\circ}$ and a declination of $\delta = $14d$\pm 19^{\circ}$. The uncertainties contain 95\% of the probability in each dimension. We then fit for the semi-major axis and semi-minor axis by minimising the loss function in Equation~\ref{eqn:lossf} using gradient descent. In Equation~\ref{eqn:lossf}, $p(\alpha',\delta')$ is the probability the event occurred at that sky location, $a$ is the semi-major axis, $b$ is the semi-minor axis, and $\lambda$ is a regularisation hyperparameter that we set to $2.7\times10^{-2}$. This regularisation corresponds to the ellipse area contributing approximately equally to the loss as the confidence interval when $|\int_{{\rm ellipse}} p(\alpha',\delta') d\alpha' d\delta'- 0.95| < 0.001$. We find that $a=13.6^{\circ}$ and $b=6.53^{\circ}$.

\begin{equation}
\label{eqn:lossf}
    \mathcal{L}(a, b) = ( \int_{{\rm ellipse}} p(\alpha',\delta') d\alpha' d\delta'- 0.95 )^2 + \lambda a b
\end{equation}


\noindent {\bf Properties of ST 200428A.} The dynamic spectrum of ST 200428A was analysed using methods previously applied to FRBs detected at the Parkes telescope$^{34}$. First, the dynamic spectra of ST 200428A at the native time and frequency resolutions (65.536\,$\mu$s and 122.07\,kHz) obtained from OVRO and GDSCC were summed. No correction for the OVRO-GDSCC geometric arrival-time delay was applied because the correction ($-7\,\mu$s) is much smaller than the time resolution. Data from Delta were not included in the sum, and in subsequent analysis, because of the increased presence of RFI at the Delta station (Extended Data Figure 1). No calibrations were applied to the data, as none were available. After excising frequency ranges at the edges of the band that are always affected by radio-frequency interference and are set to zero in the real-time pipeline (1468.232422\,MHz--1529.267578\,MHz and 1279.267578\,MHz--1280.732422\,MHz), we formed time-series in four evenly spaced sub-bands after de-dispersing at an initial DM of 332.72\,pc\,cm$^{-3}$ [16]. The time-series in each sub-band were normalised by the rms noise level in 50\,ms intervals of data on either side of the burst. As described in Ref. 34, we fit a series of models with increasing complexity, stopping when the Bayes Information Criterion (BIC) does not favour the more complex models. We found that a model combining an intrinsic width in excess of the dispersion-smeared instrumental resolution, convolved with a one-sided exponential function with a characteristic timescale scaling as $\nu^{-4}$, was negligibly preferred according to a change in BIC of 1 unit. Here, $\nu$ is the radio frequency. This corresponds to Model~3 of Ref. 34 with $\alpha=4$. The free parameters of the fit were the intrinsic width of the burst (assumed to be frequency-independent), a reference arrival time at 1529.267578\,MHz, a correction to the assumed DM, the $1/e$ timescale of the exponential function, and the burst fluences in each sub-band in data units. The convolution with the exponential is observed in several FRBs and radio pulsars, and is expected due to stochastic multi-path propagation of the burst due to refraction in inhomogeneous interstellar plasma$^{35}$. This effect is commonly referred to as ``scattering''. Despite the insignificant evidence favouring the above model over a version of this model with no scattering, we adopt it because scattering was observed in the CHIME-detected event temporally coincident with ST 200428A$^{16}$. 

The resulting best-fit burst parameters and the half-widths of their 68\% confidence intervals are given in Table~1. The $1/e$ timescale of the scattering, scaled to a frequency of 1\,GHz as is common practise in the field$^{34}$, was 0.4(1)\,ms. The model fits in each sub-band are shown in Extended Data Figure 2. The quoted band-averaged fluence (with an effective frequency of 1378\,MHz) was derived by averaging the fluences in each sub-band, and by scaling by the noise level according to the measured STARE2 system-equivalent flux density of $19\pm2$\,MJy [3], and the frequency- and time-binning. An additional scaling of 1.33 was applied to correct for the location of the burst in the STARE2 primary beam, which was negligibly different at OVRO and GDSCC. The isotropic-equivalent energy release of ST 200428A was calculated by scaling the fluence (in appropriate units) by $4\pi D^{2}\nu_{0}$, where $D$ is the distance to SGR 1935+2154 and $\nu_{0}=1374.482422$\,MHz is the midpoint of the STARE2 band. 

The adopted distance to SGR 1935+2154 is consistent with the DM of ST 200428A. According to two models for the Galactic distribution of free electrons, this DM corresponds to distances of 6--13\,kpc [36]  and 4--16\,kpc [37]. However, the significant uncertainties in these models and in the distance to SGR 1935+2154 make it difficult to numerically constrain any potential DM excess associated with the SGR. The DMs in the two models corresponding to the lowest possible distance (6.6\,kpc; Ref. 19) to SGR 1935+2154 are approximately 200\,pc\,cm$^{-3}$ [36] and 190\,pc\,cm$^{-3}$ [37]. Thus, an approximate upper limit to any potential DM excess associated with  SGR 1935+2154 is 140\,pc\,cm$^{-3}$. 

The analysis of the burst dynamic spectrum presented above provides a reference arrival time at 1529.267578\,MHz. This refers to the midpoint of the Gaussian function used to model the intrinsic burst structure. We note that it is unlikely that the burst is truly Gaussian in its intrinsic structure, and that the presence of complex structure in the time-frequency plane (as is observed in some FRBs$^{38}$) could bias this result at the approximately 0.1\,ms level. The reference arrival time was converted to a UTC arrival time at OVRO using the recorded UTC of the first spectrum in the data set. This time was converted to an arrival time at infinite frequency ($\nu=\infty$) using the fitted DM of 332.702(8)\,pc\,cm$^{-3}$, and a dispersion constant of $\frac{1}{2.41}\times10^{4}$\,s\,MHz$^{2}$\,pc$^{-1}$\,cm$^{3}$ [14]. The Earth centre arrival time was then calculated by adding a correction of 0.019210268\,s to the OVRO arrival time. This correction was derived using the known position of the OVRO station and the v2.0 astropy software package$^{39}$.

\noindent{\bf Searches for sub-threshold events from SGR 1935+2154.} We searched the STARE2 data 65.536\,ms before and after the burst for sub-threshold pulses at the same DM, and widths ranging from 0.066\,ms--4.194\,ms at each station. We found no convincing candidates. Assuming we would detect a burst with S/N$>$5, we place a limit on the fluence of other bursts in this time range of $<$400\,kJy\,ms. This fluence is obtained by scaling the measured fluence by the median S/N reported by our pipeline (18.3) divided by our sub-threshold search threshold of 5. We choose to use the median S/N reported by our pipeline because we require detections from 2 stations in order to claim that there is evidence for other pulses.

We only detect one component in ST 200428A, whereas the coincident CHIME event$^{16}$ consists of two components. The relative timing between the CHIME and STARE2 data is yet to be determined. This makes it unclear which (if any) of the two CHIME components ST 200428A corresponds to. However, assuming that ST 200428A corresponds to one of the CHIME components, the upper limit on the other CHIME component is a factor of 3.75 below the fluence of ST 200428A in the STARE2 frequency band. 

\noindent {\bf Rate calculation.} STARE2 had been observing for 448 days prior to ST 200428A. For the first 290 days, STARE2 observed only with its stations at OVRO and GDSCC. We observed with all three stations for 158 days before ST 200428A. Throughout this time, in addition to ST 200428A, we detected several solar radio bursts$^{3}$, but no other radio burst above a detection threshold of approximately 300\,kJy\,$\sqrt{{\rm ms}}$. 

To measure the all-sky rate of fast radio transients above the energy of ST 200428A, we first modelled the population of fast radio transients as a Poisson process. We then calculated our effective observing time for two epochs of observation. The first epoch corresponds to our two-station system, and the second epoch corresponds to our three-station system. We used the single-station completeness reported in Ref. 3 and an effective solid angle of 1.84 steradians to convert our observing time into an effective observing time. The single-station completeness gives the two-station system a completeness of 0.56 and the three-station system a completeness of 0.95. The solid angle was chosen such that the median S/N reported by our detection pipeline (18.3) would be at the detection threshold of 7.3 at the edge of the solid angle. This gives the area for which a burst of this fluence would have been detected. Using these parameters, we estimate a total effective observing time of 0.468 years. 

We then computed the probability of the all-sky rate for rates ranging from 0\,sky$^{-1}$\,yr$^{-1}$ to 40\,sky$^{-1}$\,yr$^{-1}$ given that we observed for 0.468 years before we found a burst. We find the all-sky rate of fast radio transients above 1.5\,MJy\,ms is 3.58$_{-2.03}^{+3.44}$\,sky$^{-1}$\,yr$^{-1}$. The reported uncertainties are 1$\sigma$ uncertainties.

To estimate the volumetric rate of this type of transient, we use the formalism developed in Ref. 3 which converts the all-sky rate of transients in a particular galaxy into a volumetric rate. The key assumption in this formalism is that the transient linearly tracks star-formation activity. This is justified by the fact that magnetars are young objects, that this object is in the plane of the Milky Way, and is associated with a supernova remnant$^{18}$. Using our computed all-sky rate, a star formation rate in the Milky Way of $1.65 \pm 0.19$\,M$_\odot$\,yr$^{-1}$ [40], and a volumetric star formation rate of $1.95\times10^{-2}$\,M$_\odot$\,Mpc$^{-3}$yr$^{-1}$ [41], we find that the volumetric rate of this type of transient is $7.23_{-6.13}^{+8.78} \times 10^7$\,Gpc$^{-3}$\,yr$^{-1}$. This is consistent with the inferred rate making an alternative assumption that transients like ST 200428A track the stellar mass of their hosts. This volumetric rate is consistent with extrapolating the luminosity function derived from a sample of bright FRBs from the Australian Square Kilometre Array Pathfinder reported in Ref. 42 down to the energy of this burst. The volumetric rate of this burst along with the FRB luminosity function is shown in Extended Data Figure 3. 

\noindent {\bf STARE2 limits on other X-ray bursts from SGR 1935+2154.} We looked through our metadata of candidate events triggered by a single station for possible missed triggers coincident with previously reported flares from SGR 1935+2154. Our candidate metadata consists of S/N ratios, arrival times, DMs, pulse widths, and number of DM, pulse width, and time trials that are consistent with being from the same candidate. We searched the metadata for candidates within one minute of reported X-ray bursts from SGR 1935+2154 and that had an elevation angle $>25^{\circ}$ at OVRO at the time of the burst. We also restricted our search to candidates with DMs between $325$\,pc\,cm$^{-3}$ and $340$\,pc\,cm$^{-3}$. We note that SGR 1935+2154 was the only SGR with known X-ray bursts that were visible to STARE2. We find no candidate events other than ST 200428A that fit this criteria at any of the three STARE2 stations.

Given our single station completeness of 0.74 [3], our completeness for this analysis is expected to be approximately 0.93 for our two station system and 0.98 for our three station system. Our three station system came online on MJD 58809, and thus only observed the X-ray bursts between MJD 58966--MJD 58968.

We used our measured SEFD, typical observing bandwidth of 188\,MHz, and threshold S/N of 7.3 to compute an upper limit on millisecond-duration radio transients for each event. We also apply a beam correction for each upper limit corresponding to the position of SGR 1935+2154 at OVRO. Our limits are shown in Extended Data Figure 4. We include ST 200428A as a blue dot. We also show our results along with the list of X-ray bursts in Extended Data Table 1.

\vspace{1cm}

\noindent {\bf Data availability statement.} Data are available upon request. These data will be placed in a public archive by the Caltech Library and provided with a DOI prior to publication of this manuscript. 

\vspace{0.5cm} 

\noindent {\bf Code availability statement.} Custom code will be made available at \\ https://github.com/cbochenek/STARE2-analysis.

\vspace{0.5cm}

\begin{enumerate}
    
    \setcounter{enumi}{32}
    \item Kawai, H. {\em et al.} Telescope Array Experiment. {\em Nucl. Phys. B} {\bf 175,} 221 (2008). 
    \item Ravi. V. The observed properties of fast radio bursts. {\em Mon. Not. R. Astron. Soc.} {\bf 482,} 1966 (2019). 
    \item Cordes, J. M. \& Chatterjee, S. Fast Radio Bursts: An Extragalactic Enigma {\em Annu. Rev. Astron. Astrophys.} {\bf 57,} 417 (2019). 
    \item Yao, J. M., Manchester, R. N. \& Wang, N. A New Electron-density Model for Estimation of Pulsar and FRB Distances. {\em Astrophys. J.} {\bf 835,} 29 (2017).
    \item Cordes, J. M. \& Lazio, T. J. W. NE2001. I. A New Model for the Galactic Distribution of Free Electrons and its Fluctuations. Preprint at \\ http://arxiv.org/abs/astroph/0207156 (2002). 
    \item Hessels, J. W. T. {\em et al.} FRB 121102 Bursts Show Complex Time-Frequency Structure. {\em Astrophys. J.} {\bf 876,} L23 (2019).
    \item Astropy Collaboration {\em et al.} The Astropy Project: Building an Open-science Project and Status of the v2.0 Core Package. {\em Astron. J.} {\bf 156,} 123 (2018). 
    \item Licquia, T. C. \& Newman, J. A. Improved Estimates of the Milky Way's Stellar Mass and Star Formation Rate from Hierarchical Bayesian Meta-Analysis. {\em Astrophys. J.} {\bf 806,} 96 (2015). 
    \item Salim, S. {\em et al.} UV Star Formation Rates in the Local Universe. {\em Astrophys. J. Suppl. Ser.} {\bf 173,} 267 (2007). 
    \item Lu, W. \& Piro, A. L. Implications from ASKAP Fast Radio Burst Statistics. {\em Astrophys. J.} {\bf 883,} 40 (2019).

\end{enumerate}

\clearpage

\begin{figure}
    \centering
    \includegraphics[width=\textwidth]{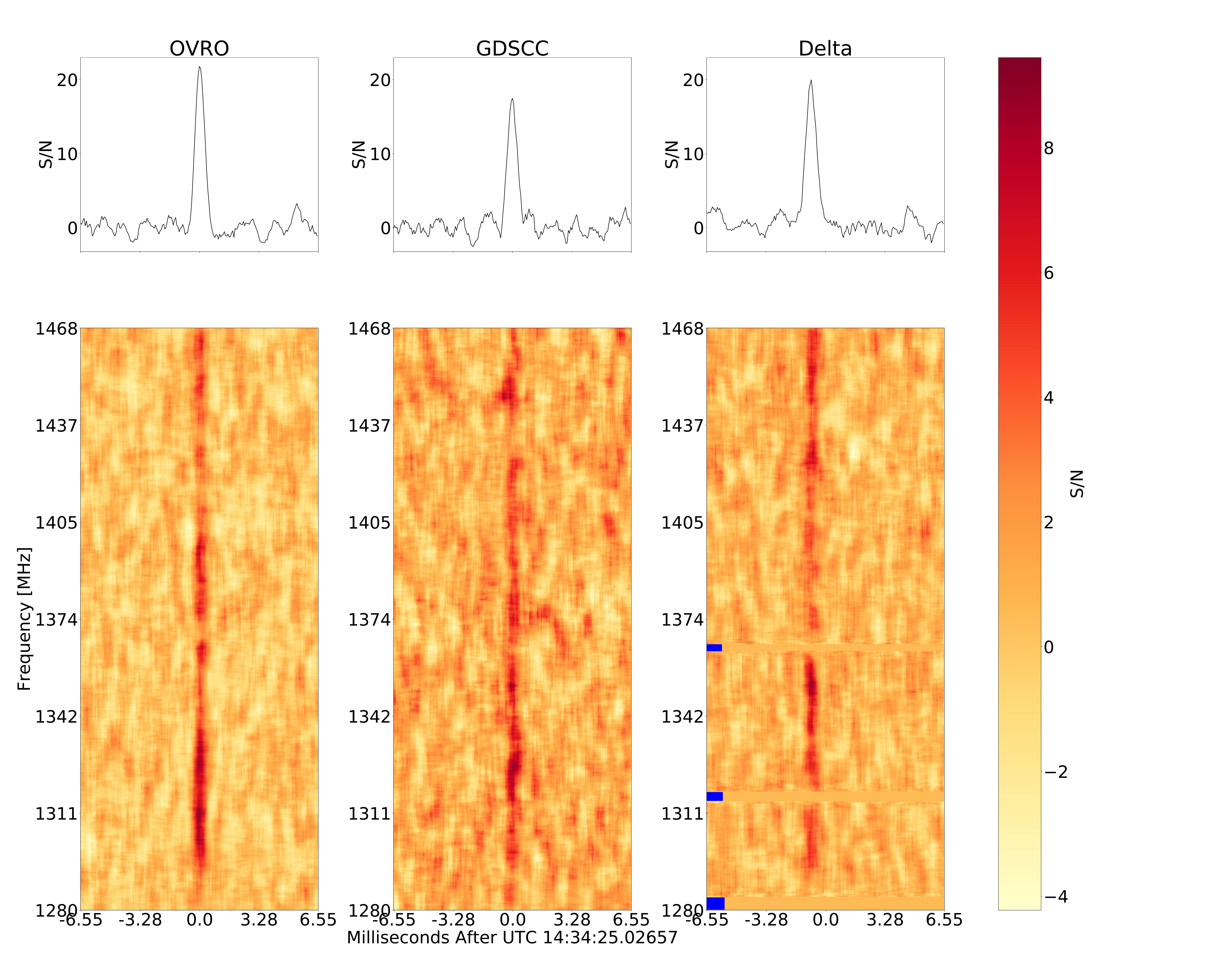}
    \caption*{{\bf Extended Data Figure 1: Time series and dynamic spectrum of ST 200428A at each station.} \textit{Top panels:} Time series at each station referenced to the arrival time at OVRO at $\nu=1529.267578$\,MHz. \textit{Bottom panels:} Dynamic spectra at each station. In all panels, the data were processed in the same way as those in Figure 1. We have not corrected for the spectral response at each station. The blue bars in the Delta dynamic spectrum indicate frequencies affected by radio-frequency interference that were excised from the data.}
    \label{fig:sites}
\end{figure}
    
    \clearpage

\begin{figure}
    \centering
    \includegraphics[width=0.75\textwidth]{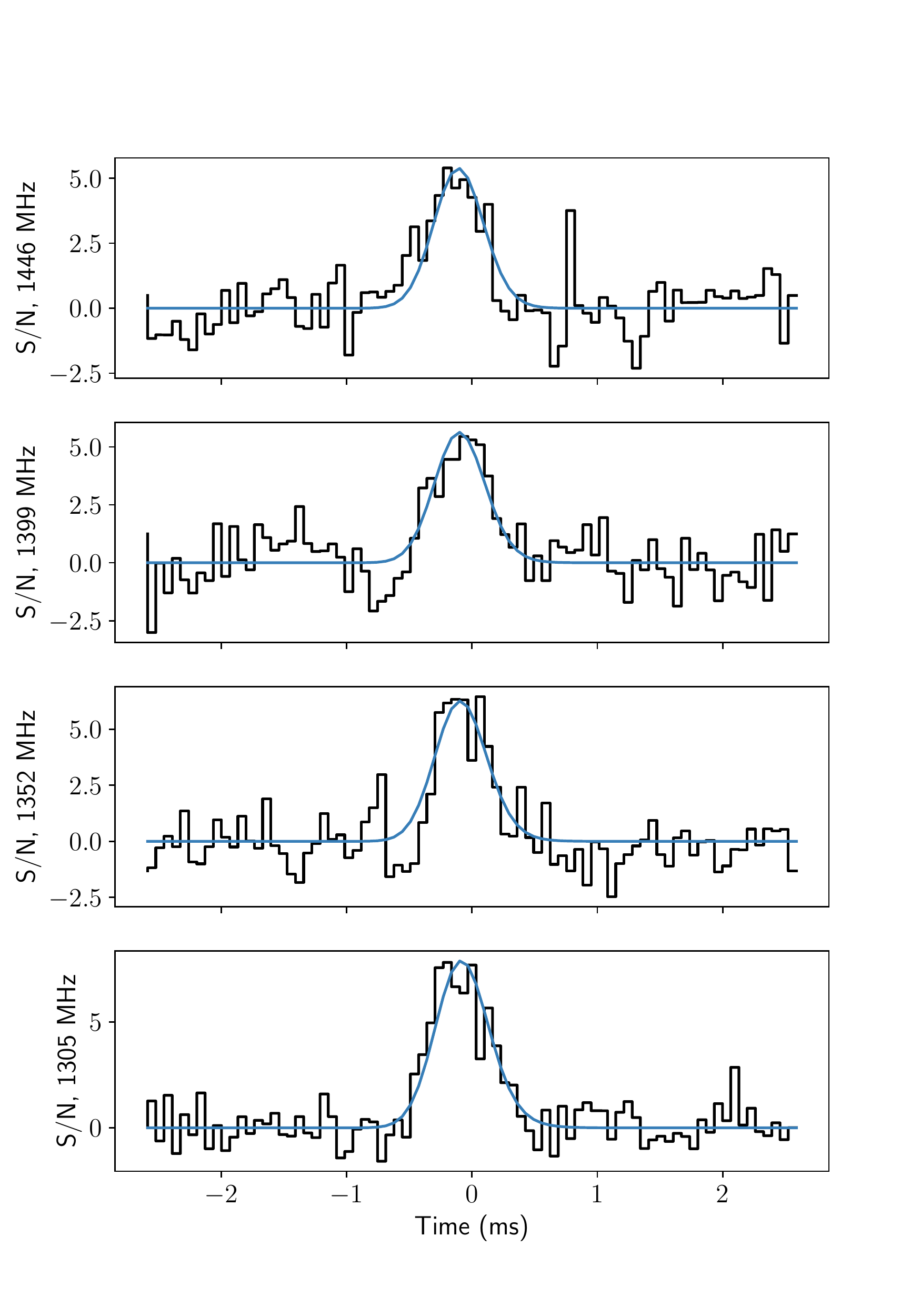}
    \caption*{{\bf Extended Data Figure 2: Fits to data on ST200428A in four sub bands.} The raw data are shown as stepped black lines, and the best-fit model is shown as smooth blue lines. The sub-band centre frequencies are indicated beside each plot.}
    \label{fig:fit}
\end{figure}

\clearpage

\begin{figure}
    \centering
    \includegraphics[width=0.9\textwidth]{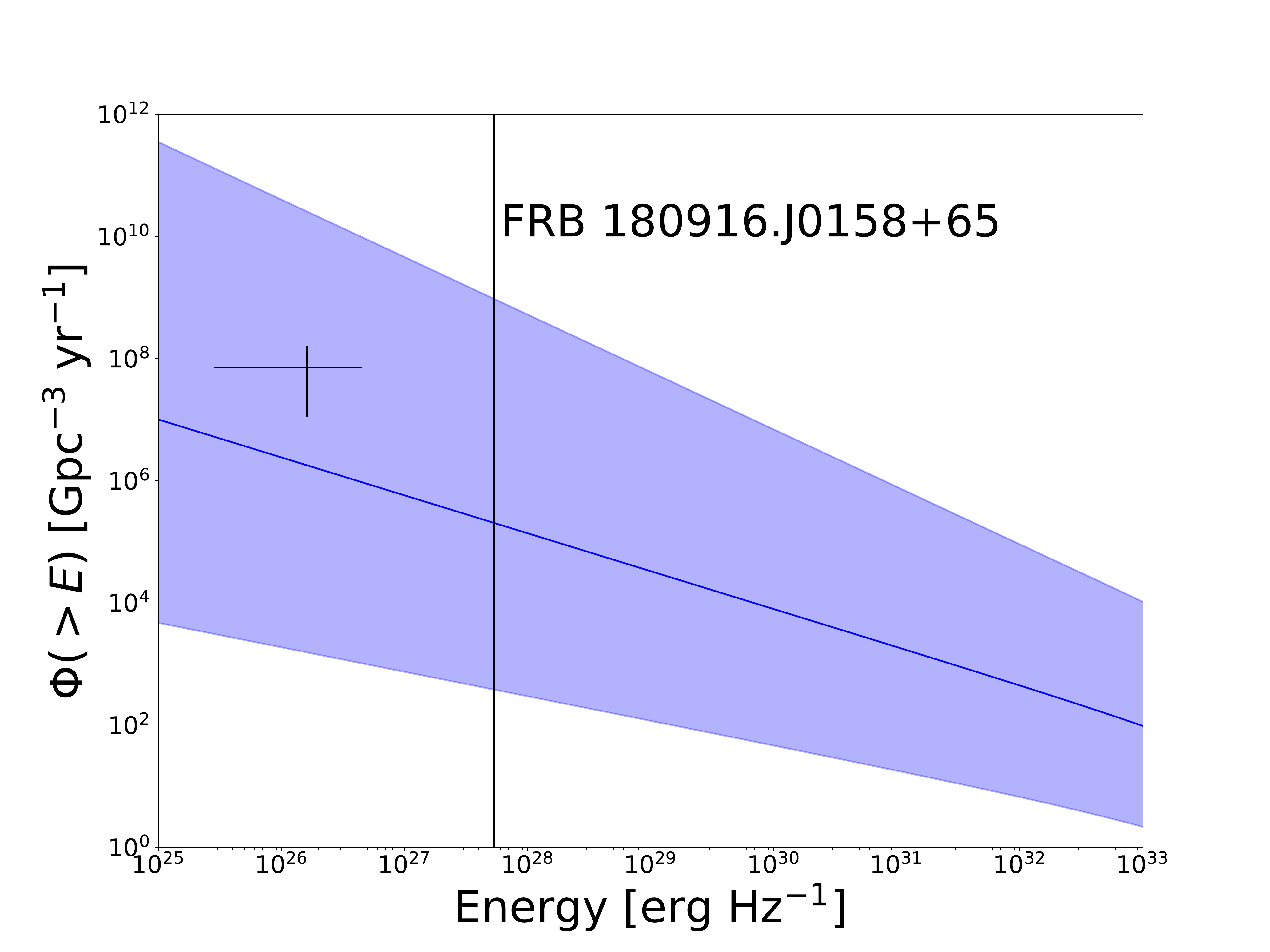}
    \caption*{{\bf Extended Data Figure 3: Volumetric rates of FRBs.} This figure shows the volumetric rate, $\Phi(>E)$, calculated from the radio burst from SGR 1935+2154 (black cross) compared with the extrapolated luminosity function of bright FRBs$^{42}$. The volumetric rate was calculated by modelling the population of Galactic fast radio transients as a Poisson process and assuming that these fast radio transients track star formation. The uncertainties on this measurement are $1\sigma$ statistical uncertainties in addition to the maximum range of possible distances to SGR 1935+2154 (4--16\,kpc)$^{37}$. We also show the energy of the weakest burst detected from FRB 180916.J0158+65, a repeating FRB, for comparison$^{4}$. The volumetric rate of Galactic fast radio transients is consistent with extrapolating the luminosity function of bright FRBs to the energy of ST 200428A. }
    \label{fig:rates}
\end{figure}

    \clearpage

\begin{figure}
    \centering
    \includegraphics[width=\textwidth]{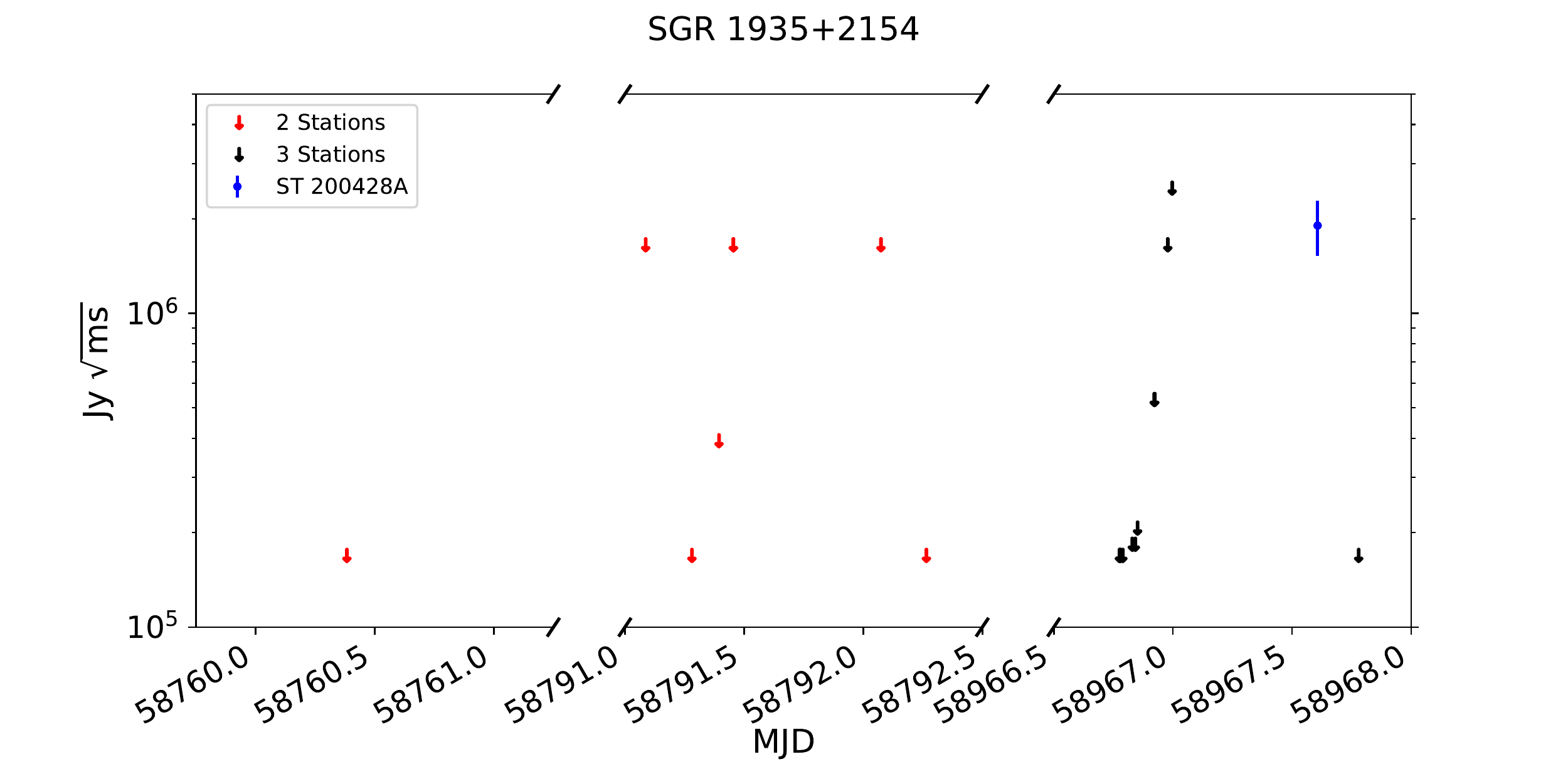}
    \caption*{{\bf Extended Data Figure 4: Upper limits on fast radio transients from other flares of SGR 1935+2154 observable by STARE2.} The ordinate shows the 7.3$\sigma$ upper limit on a potential burst's fluence in Jy\,$\sqrt{{\rm ms}}$, while the abscissa shows the MJD of reported flares from SGR 1935+2154. The derivation of the upper limits is described in the Methods section. We note that our three station system was observing only during the flares between MJD 58966 and MJD 58968, shown in black. For the other flares, only our stations at OVRO and GDSCC were observing, shown in red. We show ST 200428A in blue; the error bar represents the standard error in the measured fluence.}
    \label{fig:sgrul}
\end{figure}

\clearpage

\begin{table}
	\centering
	\caption*{{\bf Extended Data Table 1: STARE2 7.3$\sigma$ upper limits on reported X-ray bursts from SGR 1935+2154 that occurred in the STARE2 field of view.}  \\
	$^{\rm a}$ The upper limits represent a threshold S/N of 7.3$\sigma$. \\
	$^{\rm b}$ ``GCN'' refers to the GRB Circular Network and ``ATel'' refers to the Astronomers Telegram.}
	\label{tab:1}
	\begin{tabular}{cccc} 
		\hline
		 {\bf MJD} & {\bf OVRO Elevation} & { \bf Limit (MJy $\sqrt{\rm \mathbf{ms}}$)$^{\rm a}$} & {\bf Citation$^{\rm b}$} \\
		\hline
58760.3756215 & 73$^{\circ}$ & 0.17 & GCN 25975 \\
58791.0795949 & 31$^{\circ}$ & 1.65 & GCN 26169 \\
58791.2736111 & 74$^{\circ}$ & 0.17 & GCN 26153 \\
58791.387419 & 49$^{\circ}$ & 0.39 & GCN 26160 \\
58791.4475231 & 32$^{\circ}$ & 1.65 & GCN 26160 \\
58791.4475302 & 32$^{\circ}$ & 1.65 & GCN 26242 \\
58792.066956 & 29$^{\circ}$ & 1.65 & GCN 26171 \\
58792.2577431 & 75$^{\circ}$ & 0.17 & GCN 26242 \\
58966.768287 & 74$^{\circ}$ & 0.17 & ATel 13675, GCN 27659 \\
58966.7722222 & 74$^{\circ}$ & 0.17 & GCN 27663 \\
58966.7729051 & 74$^{\circ}$ & 0.17 & GCN 27664 \\
58966.7820486 & 75$^{\circ}$ & 0.17 & GCN 27663 \\
58966.8220411 & 71$^{\circ}$ & 0.18 & GCN 27667 \\
58966.8345602 & 68$^{\circ}$ & 0.18 & GCN 27661 \\
58966.8439815 & 66$^{\circ}$ & 0.21 & GCN 27664 \\
58966.9139236 & 47$^{\circ}$ & 0.53 & GCN 27661 \\
58966.9162289 & 46$^{\circ}$ & 0.53 & GCN 27667 \\
58966.9708333 & 30$^{\circ}$ & 1.65 & GCN 27663 \\
58966.9892552 & 25$^{\circ}$ & 2.51 & GCN 27667 \\
58967.771956 & 75$^{\circ}$ & 0.17 & GCN 27667 \\

		\hline
	\end{tabular}
\end{table}

\end{document}